\documentclass[iop]{emulateapj}
\usepackage{color}
\usepackage{graphicx}
\usepackage{amsmath}

\shorttitle{Testing CDDR with high-$z$ quasars} \shortauthors{Zheng,
et al.}

\begin{document}
\title{Multiple measurements of quasars acing as standard probes: exploring the cosmic distance duality relation at higher redshift}
\author{Xiaogang Zheng\altaffilmark{1}, Kai Liao\altaffilmark{2}, Marek
Biesiada\altaffilmark{3}, Shuo Cao\altaffilmark{3$\ast$}, Tong-Hua
Liu\altaffilmark{3}, Zong-Hong Zhu\altaffilmark{1,3$\dag$} }

\altaffiltext{1}{School of Physics and Technology, Wuhan University,
Wuhan 430072, China;} \altaffiltext{2}{School of Science, Wuhan
University of Technology, Wuhan 430070,
China;}\altaffiltext{3}{Department of Astronomy, Beijing Normal
University, Beijing 100875, China; \emph{caoshuo@bnu.edu.cn;
zhuzh@bnu.edu.cn}}

\begin{abstract}

General relativity reproduces main current cosmological
observations, assuming the validity of cosmic distance duality
relation (CDDR) at all scales and epochs. However, CDDR is poorly
tested in the redshift interval between the farthest observed Type
Ia supernovae (SN Ia) and that of the Cosmic Microwave background
(CMB). We present a new idea of testing the validity of CDDR,
through the multiple measurements of high-redshift quasars.
Luminosity distances are derived from the relation between the UV
and X-ray luminosities of quasars, while angular diameter distances
are obtained from the compact structure in radio quasars. This will
create a valuable opportunity where two different cosmological
distances from the same kind of objects at high redshifts are
compared. Our constraints are more stringent than other currently
available results based on different observational data and show no
evidence for the deviation from CDDR at $z\sim 3$. Such accurate
model-independent test of fundamental cosmological principles can
become a milestone in precision cosmology.

\end{abstract}


\maketitle

\section{Introduction}

As a fundamental relation rooted in the very ground of modern
cosmology, i.e., the validity of General Relativity (or more general
-- the metric theory of gravity), the so-called cosmic distance
duality relation (CDDR) is very successful in explaining many
observational facts concerning our Universe including large-scale
distribution of galaxies and the near-uniformity of the CMB
temperature \citep{Planck18}. More specifically, there are two basic
measurable distances useful in cosmology: the angular diameter
distance $D_A(z)$ and the luminosity distance $D_L(z)$. They are
related to each other according to $D_A(z)(1+z)^2/D_L(z)=1$
\citep{Etherington33, Etherington07}. This relation holds under
three very general assumptions: metric theory of gravity, assumption
that photons travel along null geodesics and assumption that the
number of photons is conserved in the beam. Any deviation from this
relation would signal violation of these assumptions, i.e. the new
physics or that photon number is not conserved, most likely because
of the impact of intergalactic medium \citep{Liao15,Qi19a}.

Different methods have been used to test the validity of CDDR
\citep{Cao11a, Cao11b, Li11, Costa15, Liao16, Holanda17, Rana17}.
All these methods depend on the measurements of angular diameter
distance and luminosity distance to cosmological sources and each of
them have their own advantages or drawbacks according to the objects
observed. Most of the CDDR tests performed so far were based on low
redshift objects. For example, \citep{Holanda10, Li11, Yang13}
combined angular diameter distances from galaxy clusters and
luminosity distances from type Ia supernovae. Due to limitations
concerning availability of appropriate observational data it has
been hard to get distances (especially angular diameter distances)
from high redshift objects. \citet{Liao16} proposed a new method to
test the CDDR with strong lensing systems in which the angular
diameter distance ratio $R_A=D_{ls}/D_{s}$ can be measured from
image separations, provided the stellar central velocity dispersion
$\sigma_0$ is measured. In the above formula, $D_{ls}$ and $D_s$
denote angular diameter distances from the source to lens and from
the source to observer, respectively. Benefiting from the higher
redshift attainable in strong gravitational lensing systems
(comparing to galaxy clusters), this method was extended to
Gamma-Ray Burst measurements \citep{Holanda17} and strong
gravitational lensing time delay measurements \citep{Rana17}.
Moreover, the possibility of using the angular size-redshift
relation from compact radio sources
\citep{Kellermann93,Gurvits94,Jackson06, Cao15,Cao17a} to get
angular diameter distances has recently attracted attention. In
particular, \citet{LiX18} combined the ultra-compact radio sources
with type Ia supernova to test the validity of CDDR.

From the perspective of CDDR test it would be advantageous to have
objects spanning a wide redshift range with both angular diameter
and luminosity distances measurable. A lot of attempts have been
made to explore whether quasars can be such a kind of probes. Their
luminosity distances were proposed to be assessed from the relation
between the broad line region (BLR) radius of the reverberation
mapping (RM) method and monochromatic luminosity \citep{Watson11},
the properties of super-Eddington accreting massive black holes
\citep{Wang13}, the non-linear relation between the ultraviolet (UV)
and X-ray luminosity \citep{Risaliti15}.  On the other hand, angular
diameter distances to the quasars could be derived from the
classical geometrical size of the BLR \citep{Elvis02} or the angular
size - redshift relation of compact structures in
intermediate-luminosity radio quasars  \citep{Cao17a}. Although it
is hard to test the CDDR on individual quasars having both
luminosity and angular diameter distance measured, yet it is
tempting to use already existing rich statistical material of
objects falling into separate classes: standardizable candles and
rulers. In this paper we will use luminosity distances to quasars
from the non-linear relation between the UV and X-ray emission
\citep{Risaliti19} and angular diameter distances from the angular
size - redshift relation of compact radio sources \citep{Cao17a}.

In Section \ref{sec:data}, we briefly introduce methodology of
deriving two different cosmological distances from quasar
measurements in X-ray, UV and radio bands. In order to discuss the
influence of cosmological model on the theoretical angular sizes and
the sensitivity of Gaussian processes to the choice of mean
function, we assume the Hubble constant as $\rm
H_0=70\;km\;s^{-1}\;Mpc^{-1}$. In Section \ref{sec:result}, we
present our results and discuss the advantages and disadvantages of
our statistical method. Finally, we summarize our conclusions in
Section \ref{sec:con}.

\section{Data and Methodology}\label{sec:data}

\subsection{Luminosity distances from a nonlinear X-UV luminosity relation of quasars}\label{subsec:QSO_XUV}

Benefitting from high luminosity and large redshift coverage,
quasars are very  promising cosmological probes. One can assess
their luminosty distances indirectly from the correlations between
spectral features and luminosities
\citep{Watson11,Wang13,LaFranca14}. Quite recently, the nonlinear
relation between the UV and X-ray luminosities of quasars has been
used to construct the ``Hubble Diagram" \citep{Risaliti15,Lusso16,
Risaliti17, Bisogni18, Risaliti19} and test the cosmological model.
From now on we will use the abbreviation QSO[XUV] to denote this
method. The nonlinear relation between the UV and X-ray luminosities
can be expressed as $\log(L_X)=\gamma \log(L_{UV})+\beta$ where
$\gamma$ is the slope and $\beta$ is the intercept. According to the
flux-luminosity relation of $F=L/4\pi D^2_L$, it can be rewritten as
a relation between the observed fluxes:
\begin{equation}\label{eq:FXFUV}
\log F_X=\gamma \log F_{UV}+2(\gamma-1)\log D_L+(\gamma-1) \log
(4\pi)+\beta
\end{equation}
In order to be useful for cosmological inference, the $L_X$-$L_{UV}$
relation should not evolve with redshift, i.e. the slope $\gamma$
and the intercept $\beta$ should be constant. While one is able test
the redshift evolution of the slope on quasar sub-samples in
different redshift intervals, it is still hard to check the
intercept before the solid physical understanding of the
$L_X$-$L_{UV}$ relation is gained (for details see
\citet{Risaliti15}). Another issue worth consideration is the global
intrinsic dispersion $\delta$ which is much larger than typical
uncertainties of observational fluxes.

\begin{figure}[htbp]
\begin{center}
\centering
\includegraphics[angle=0,width=75mm]{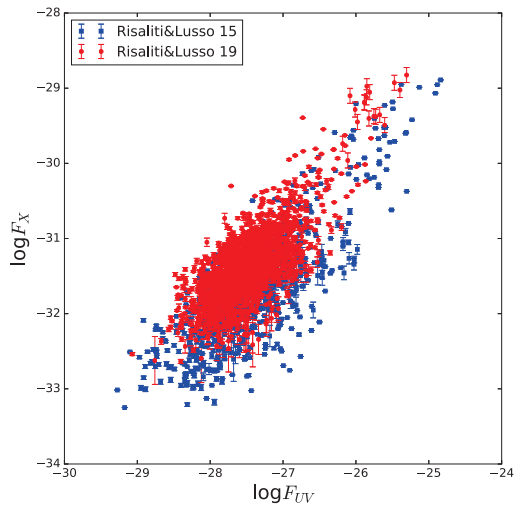}
\caption{\label{figFXFUV} $\log F_{X}$ vs. $\log F_{UV}$ diagram of
quasars. Blue squares represent 808 data points from
\citep{Risaliti15}, while the red circles represent 1598 data points
from \citep{Risaliti19}.}
\end{center}
\end{figure}

After discarding the broad absorption line and radio-loud quasars
which obviously deviate from the $L_X$-$L_{UV}$ relation and
considering the influence of dust extinction, \citet{Risaliti15}
extracted the ``best sample" of 808 quasars suitable for
cosmological inference. Analyzing the redshift dependence of the
slope they derived the average value of $\gamma=0.6\pm0.02$ and
dispersion $\delta=0.3$. Most recently, \citet{Risaliti19} collected
a sample of 7238 quasars with available X-ray and UV measurements,
and selected 1598 quasars suitable for cosmological applications.
Richer statistics resulted in more accurate slope determination
$\gamma=0.633\pm0.002$ and smaller dispersion $\delta=0.24$.  X-ray
and UV fluxes of the above mentioned two samples are shown for
comparison in Fig.~\ref{figFXFUV}. Both of them have large intrinsic
dispersions, the slopes seem similar but the intercept displays
deviating  tendency. While using these measurements to perform the
Markov chain Monte Carlo (MCMC) estimation of cosmological model
parameters, one should fix the slope $\gamma$ at the value which was
estimated from narrow redshift bins or treat it as a free parameter
to provide consistent results \citep{Risaliti15, Risaliti19,
Melia19}. The intercept $\beta$, however, needs to be
cross-calibrated. We have used the most recent compilation from
\citet{Risaliti19}. As discussed in more details in Section
\ref{subsec:CDDR_Test}, we fixed the slope $\gamma$ and treated the
intercept $\beta$ as a nuisance parameter. The dispersion $\delta$
was also treated as a free parameter contributing to the intrinsic
scatter.

\begin{figure}[htbp]
\begin{center}
\centering
\includegraphics[angle=0,width=75mm]{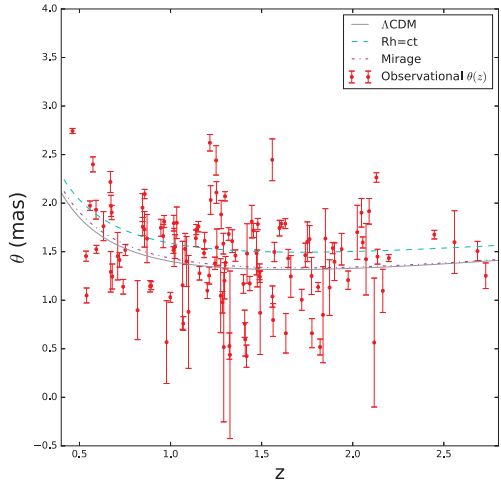}
\caption{\label{figtheta} Angular size $\theta$ versus redshift $z$
for the compact structure in 120 intermediate-luminosity quasars.
Grey solid line, cyan dashed line and magenta dash-dot line
illustrate theoretical angular size vs. redshift relation calculated
from the fiducial $\Lambda$CDM cosmology with $\Omega_m=0.27$, the
so called $\rm R_h=ct$ cosmology and the Mirage cosmology
($\Omega_m=0.27$, $w_0=-0.7$, $w_1=-1.09$), respectively. The linear
size of $l_m=11.42\;\text{pc}$ calibrated with SN Ia in
\cite{Cao17b} was used to calculate the theoretical angular size. }
\end{center}
\end{figure}

\begin{figure}[htbp]
\begin{center}
\centering
\includegraphics[angle=0,width=75mm]{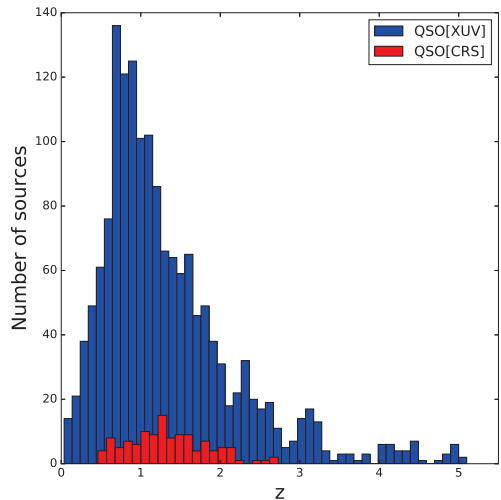}
\caption{\label{figz} The comparison of redshift distribution of the
samples studied. Blue and red histograms represent redshift
distributions of QSO[XUV] and QSO[CRS], respectively.}
\end{center}
\end{figure}

\begin{figure}[htbp]
\begin{center}
\centering
\includegraphics[angle=0,width=75mm]{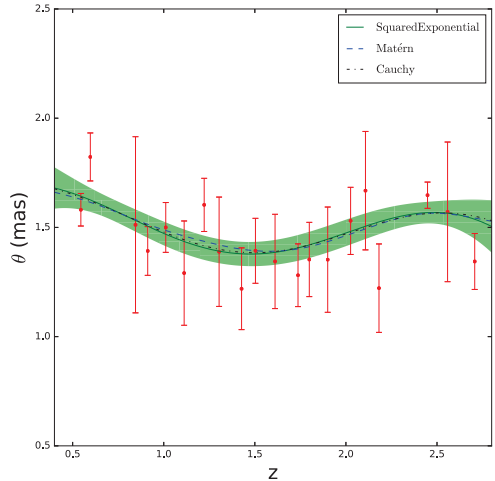}
\caption{\label{figtheta2} Angular size vs. redshift relation
reconstructed from binned data with different covariance functions
and zero mean function. Red points represent median values of 20
redshift bins with width $\Delta z=0.1$ starting from the smallest
redshift of the intermediate-luminosity quasars. Green solid line,
blue dashed line and black dash-dot line means the reconstructed
angular size with Squared Exponential, $\rm Mat\acute{e}rn$ and
Cauchy covariance function, respectively. Green range means the
$1\sigma$ uncertainty band of the reconstructed angular size based
on the Squared Exponential convariance function.}
\end{center}
\end{figure}

\subsection{Angular diameter distances from compact radio quasars}\label{subsec:QSO_CRS}

The angular size-redshift ($\theta-z$) relation of the compact
structure in radio sources has been useful for cosmological studies.
The angular size of  compact sources at different redshifts
$\theta(z)$ can be written as
\begin{equation}\label{eq:thetaz}
\theta(z)=\frac{l_m}{D_A(z)}
\end{equation}
where $l_m$ is the intrinsic metric length of the source and
$D_A(z)$ is the angular diameter distance. The angular size-redshift
relation was first proposed by \cite{Kellermann93} who attempted to
estimate the deceleration parameter based on 79 compact sources
obtained using Very Long Baseline Interferometry (VLBI) at 5 GHz
frequency where the angular size was defined as a distance between
the core and the component of $2\%$ core brightness. The method was
extended by \cite{Gurvits94} who used 337 active galactic muclei
(AGNs) observed at 2.29 GHz by \cite{Preston85} to discuss the
luminosity and redshift dependence of their characteristic size and
estimate the deceleration parameter. Unlike the angular size
definition of \cite{Kellermann93}, \cite{Gurvits94} used the modulus
of visibility $\Gamma=S_c/S_t$ to express source compactness and the
characteristic angular size of radio sources can be calculated
through the expression of $\theta(z)=2\sqrt{-\ln\Gamma \ln 2}/\pi
B_{\theta}$ where $B_{\theta}$ is the interferometer baseline, $S_c$
and $S_t$ are correlated flux density and total flux density,
respectively. Moreover, \citet{Gurvits99} used another sample of 330
VLBI contour maps at 5 GHz collected from the literature and
discussed the influence of dispersion in the ``$\theta-z$" relation
and these data was widely used to cosmological parameter inference
\citep{Chen03,Zhu04}. Then, \citet{Jackson04} tried to establish a
plausible model to understand the physical meaning of such standard
rulers using the compilations from \citet{Gurvits94,Gurvits99}.

Phenomenological dependence of the intrinsic length in
Eq.~(\ref{eq:thetaz}) on the source luminosity and the redshift can
be expressed as
\begin{equation}\label{eq:lm}
l_m=lL^{\rho}(1+z)^n
\end{equation}
where $l$ is the linear size scaling factor, $\rho$ and $n$
power-law exponents capture the dependence of the intrinsic length
on source luminosity and redshift, respectively. In
\cite{Preston85}, 917 radio sources were detected in several VLBI
studies at 2.29 GHz according to 1398 candidates selected from
previous surveys while some of them did not have the necessary
information like the total flux density or redshift. Then,
\cite{Jackson06} complemented these data with relevant information
obtained from the NASA/IPAC Extragalactic Database and
contemporaneous radio measurements catalogue. The resulting
compilation comprised 613 object in total. However, it was a mixture
of extragalactic objects including quasars, BL Lacertae objects,
radio galaxies, etc. As discussed in \cite{Cao15}, the dependence of
the  linear size on luminosity and redshift is different in
different class of objects. This conclusion was based on the
\cite{Gurvits94} sample under assumption of standard $\Lambda$CDM
cosmological model and with best fitted parameters obtained from
Planck/WMAP observations. Applying the selection criteria of flat
spectral index $-0.38\leq\alpha\leq0.18$ and luminosity in the range
of $10^{27}\; \text{W/Hz}\leq L \leq 10^{28}\; \text{W/Hz}$ to the
\cite{Jackson06} sample, \cite{Cao17a, Cao17b} identified a sample
of 120 intermediate-luminosity quasars displaying negligible
dependence on both source luminosity and redshift
($|\rho|\approx10^{-4}, |n|\approx10^{-3}$). Therefore the compact
structure sizes of these quasars are potentially promising standard
rulers with multi-frequency VLBI observations \citep{Cao18}. This
subsample was successfully used in cosmological applications
including the measurements of speed of light \citep{Cao17b,Cao20}
and cosmic curvature at different redshifts \citep{Cao19},
cosmological model selection \citep{Li17,Ma17}, testing modified
gravity models \citep{Qi17, Xu18} and interacting dark energy models
\citep{Zheng17}. We will use it in the present paper and denote it
by the abbreviation QSO[CRS]. Fig.~\ref{figtheta} displays the
angular size-redshift relation in this sample (red points with
corresponding error bars).

\subsection{The test of CDDR} \label{subsec:CDDR_Test}

The validity of CDDR can be tested by the determination of the
$\eta(z)$ parameter:
\begin{equation}\label{eq:DDR}
\eta(z)=\frac{D_L(z)}{D_A(z)(1+z)^2}
\end{equation}
where $\eta(z)=1$ corresponds to the standard Etherington
reciprocity relation and any statistically significant deviations
from it might signal the violation of any of the underlying
assumptions. In our study we derived luminosity distances and
angular diameter distances   from QSO[XUV] and QSO[CRS],
respectively. The advantage of this approach is in covering the wide
redshift range and using the same population of objects (the
quasars) in the assessment of two different distance measures. As
discussed above, we used the compilation of angular size
measurements of compact radio quasar to determine the angular
diameter distance, and the quasar flux measurements in X-ray and UV
bands which can be used to derive the luminosity distance. Redshift
distributions of QSO[XUV] and QSO[CRS] samples are shown in
Fig.~\ref{figz}. One can see that the redshift range probed by
quasars is considerable and that two samples overlap sufficiently
with each other.

In order to test the CDDR, we use four parameterizations of
$\eta(z)$
\begin{equation}\label{eq:DDReta}
\eta(z)=\left\{\begin{array}{l@{\hspace{1cm}}l}
1+\eta_{0}\;z \\
(1+z)^{\eta_1} \\
1+\eta_{2}\;\frac{z}{1+z} \\
1+\eta_{3}\;{\rm ln}(1+z) \\
\end{array}\right.
\end{equation}
There is no guidance from the theory which parametrization could be
distinguished, but mutually consistent results for different
parameterizations would strengthen the robustness of the conclusion.

From Eq.~(\ref{eq:FXFUV}), theoretical expression of the luminosity
distance can be reformulated as
\begin{eqnarray}\label{eq:DLxuv}
\nonumber D_L^{QSO[XUV]}(z)&=&10^{(\log F_X-\gamma \log
F_{UV})/(2\gamma-2)}\\
&\times&10^{-[\log(4\pi)/2+\beta/(2\gamma-2)]}
\end{eqnarray}
In principle, the parameters $\gamma$ and $\beta$ can be fitted to
different cosmological models in a manner analogous to the Type Ia
Supernova light-curve fitting \citep{Betoule14, Melia19}. Other
approach is to derive the slope $\gamma$ from the nonlinear relation
between the UV and X-ray luminosities with sub-samples in different
redshift intervals and calibrate the $\beta$ intercept on external
probes like SN Ia \citep{Risaliti15}. \cite{Risaliti15} demonstrated
that the   $\gamma$ parameter does not display any significant
redshift evolution pattern but is close to a certain average value.
While it is not possible to test the redshift dependence of the
intercept parameter $\beta$, one can anchor it by means of the
cross-calibration on the SN Ia sample having redshifts overlapping
with quasars. Following this method, we assumed $\gamma = 0.633 \pm
0.002$ derived in \citet{Risaliti19} but we treated the intercept
$\beta$ as a free parameter.

Similarly, according to Eq.~(\ref{eq:thetaz}), angular diameter
distances can be calculated from observed angular sizes as
\begin{equation}\label{eq:DAcrs}
D_A^{QSO[CRS]}(z)=l_m/\theta(z)
\end{equation}
and we also treated \textbf{the length scale $l_m$ } as a free
parameter. In order to test the CDDR using different samples one
should use a redshift matching criterion $\Delta z<0.005$
\citep{Li11,Liao16}. However, it turned out difficult to fulfill
this criterion in order to compare distances derived from QSO[XUV]
and QSO[CRS] directly. Therefore, proceeding in a similar manner as
\citet{LiX18}, we  reconstruct QSO[CRS] angular size as a function
of redshift from the binned data. The angular size of
intermediate-luminosity quasars was grouped into 20 redshift bins of
width $\Delta z=0.1$ starting from the smallest redshift of this
sample. Median values of angular size plotted against the mean
redshift in each bin are shown in Fig.\ref{figtheta2}. The python
package GaPP based on Gaussian Processes \citep{Seikel12} was used
for the reconstruction process which depends on the mean function
and the covariance function $k(x,\tilde{x})$. In order to discuss
the influence of the choice of these two prior functions, we studied
four mean functions and three covariance functions. The prior mean
functions which we discussed are the following: zero mean function,
the theoretical function of angular size calculated from the angular
diameter distance under the assumption of three cosmological models:
flat $\Lambda$CDM with $\Omega_m=0.27$, so called $R_h=ct$ Universe
\citep{Melia12}, and a Mirage model with $\Omega_m=0.27$, $w_0=-0.7$
and $w_1=-1.09$ \citep{Shafieloo12}. The linear size scaling factor
$l_m=11.42\;\rm pc$ was assumed as calibrated with SN Ia in
\cite{Cao17b}. These functions (except the zero mean function) are
shown in Fig.\ref{figtheta}. There are many possible covariance
functions and we studied three most popular ones. They comprise: the
Squared Exponential function
\begin{equation}
k(x,\tilde{x})=\sigma_f^2\exp\left(-\frac{(x-\tilde{x})^2}{2\ell^2}\right)
\end{equation}
the $\rm Mat\acute{e}rn$
\begin{equation}
k(x,\tilde{x})=\sigma_f^2\exp\left[-\frac{\sqrt{3}|x-\tilde{x}|}{\ell}\right]\left(1+\frac{\sqrt{3}|x-\tilde{x}|}{\ell}\right)
\end{equation}
and Cauchy covariance function
\begin{equation}
k(x,\tilde{x})=\sigma_f^2\frac{\ell}{(x-\tilde{x})^2+\ell^2}
\end{equation}
where $\sigma_f$ and $\ell$ are hyperparameters that control the
amplitude of deviation from the mean function and the typical length
scale in x-direction, respectively. It is instructive to discuss the
effects of the mean function and covariance function selection
(prior assumptions) on the reconstruction. In order to show the
impact of covariance function choice, we fixed the zero mean and
preformed reconstruction with three different covariance functions
mentioned above. The result obtained under assumption of Squared
Exponential covariance function is illustrated in
Fig.~\ref{figtheta2}, where the green solid line represents the
reconstructed $\theta(z)$ relation and green region around it
represents $1\sigma$ uncertainty band. The blue dashed line and
black dash-dot line represent the reconstructed $\theta(z)$ relation
with the Mat$\rm\acute{e}$rn and Cauchy covariance functions,
respectively. Their uncertainty bands are not shown in order not to
blur the picture since they are similar to the one displayed. One
can see that differences between reconstructions performed with
different choice of covariance function are insignificant.
Similarly, we checked sensitivity of reconstructions with respect to
the choice of the mean function fixing the covariance as Squared
Exponential one and using three main functions mentioned above. It
turned out that the impact of mean function choice on the
reconstruction was even smaller that that of the covariance
function. Therefore for further calculations we assumed the zero
mean function and the Squared Exponential covariance function to get
the reconstructed $\theta(z)$ function (i.e. the green line and
region in Fig.\ref{figtheta2}). Using this reconstructed relation we
were able to have a one-to-one matching between the QSO[CRS] angular
diameter distance and the QSO[XUV] luminosity distance at the same
redshift.

\begin{figure*}[htbp]
\begin{center}
\centering
\includegraphics[angle=0,width=70mm]{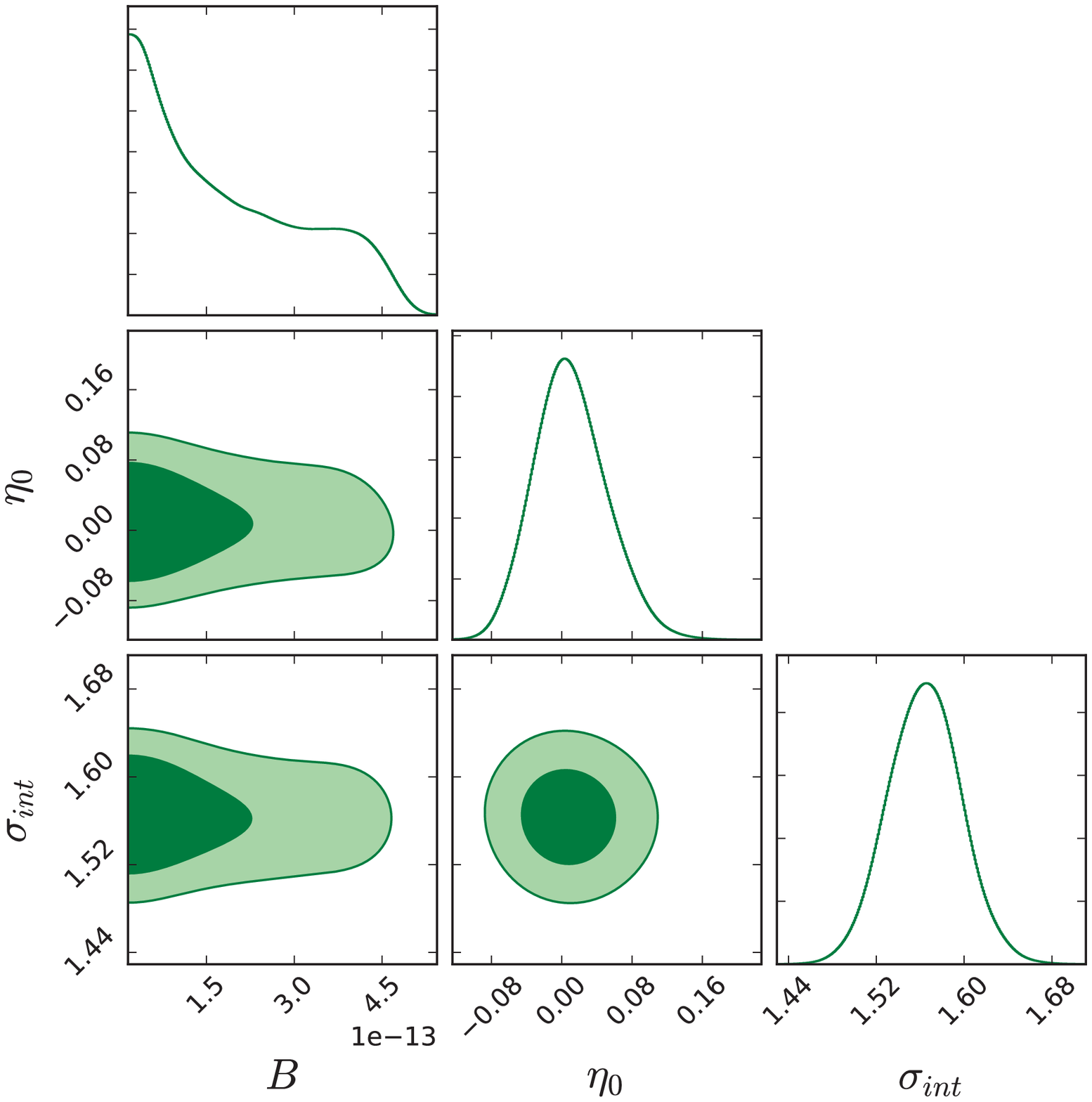}
\includegraphics[angle=0,width=70mm]{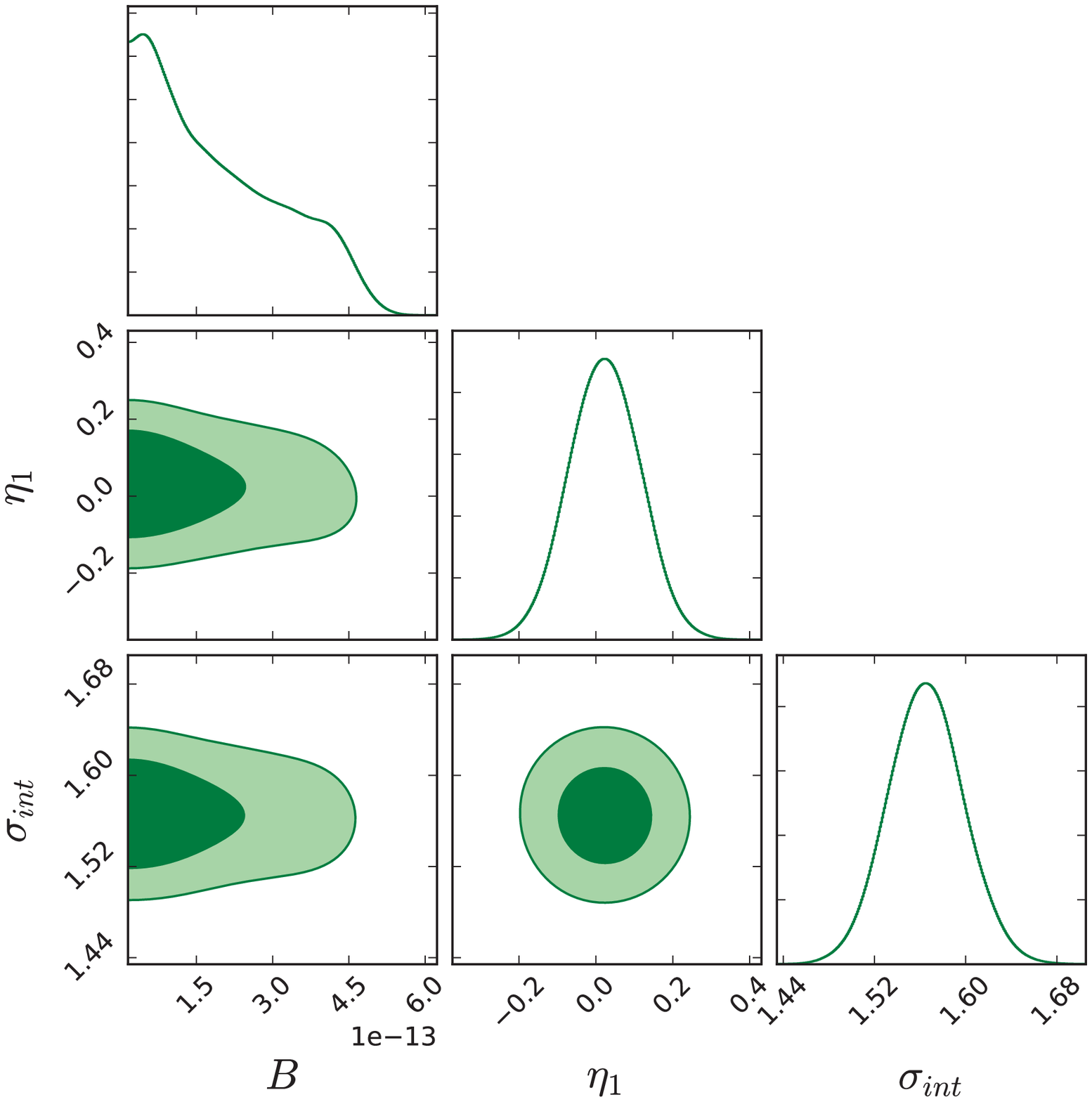}
\includegraphics[angle=0,width=70mm]{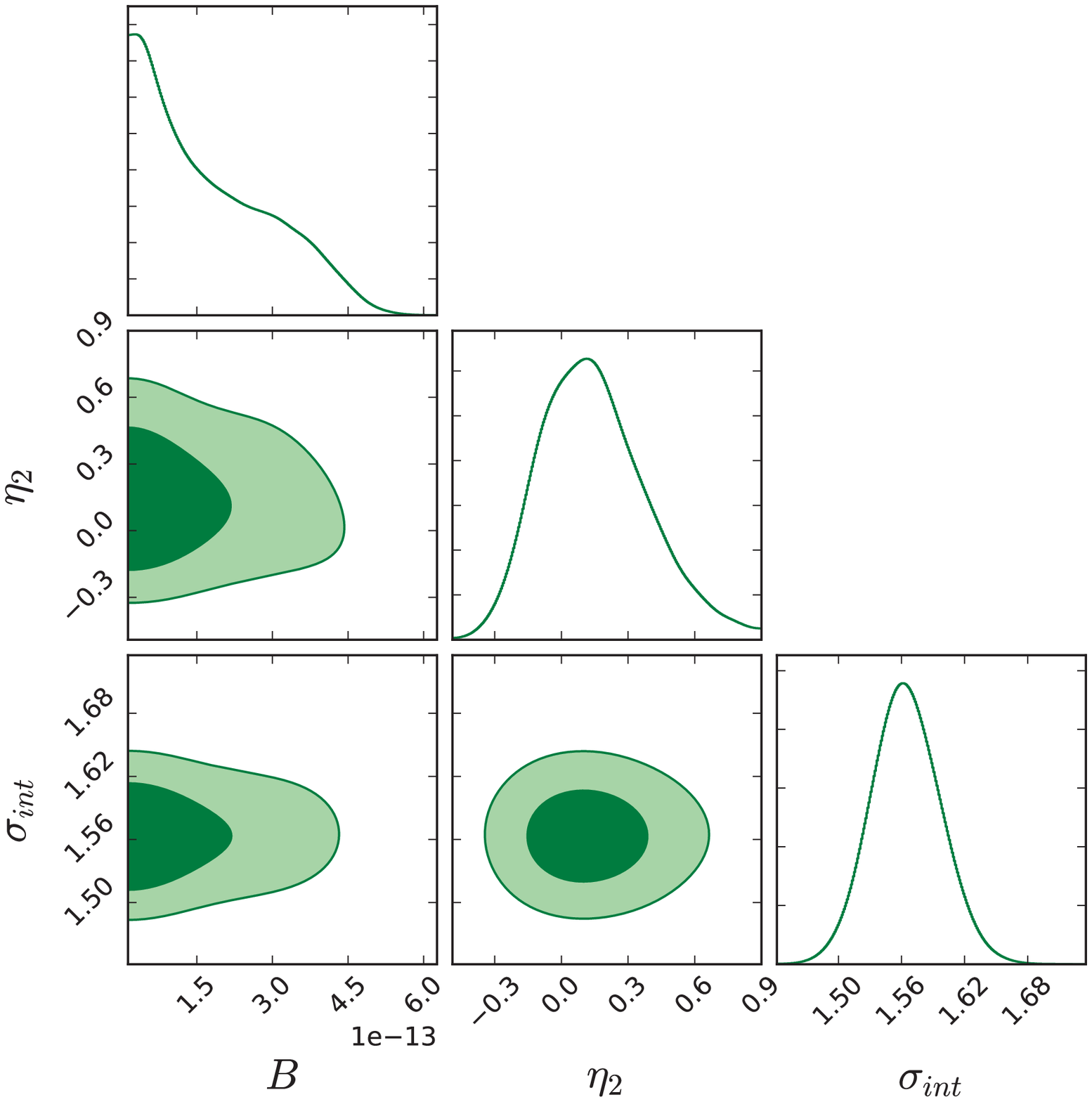}
\includegraphics[angle=0,width=70mm]{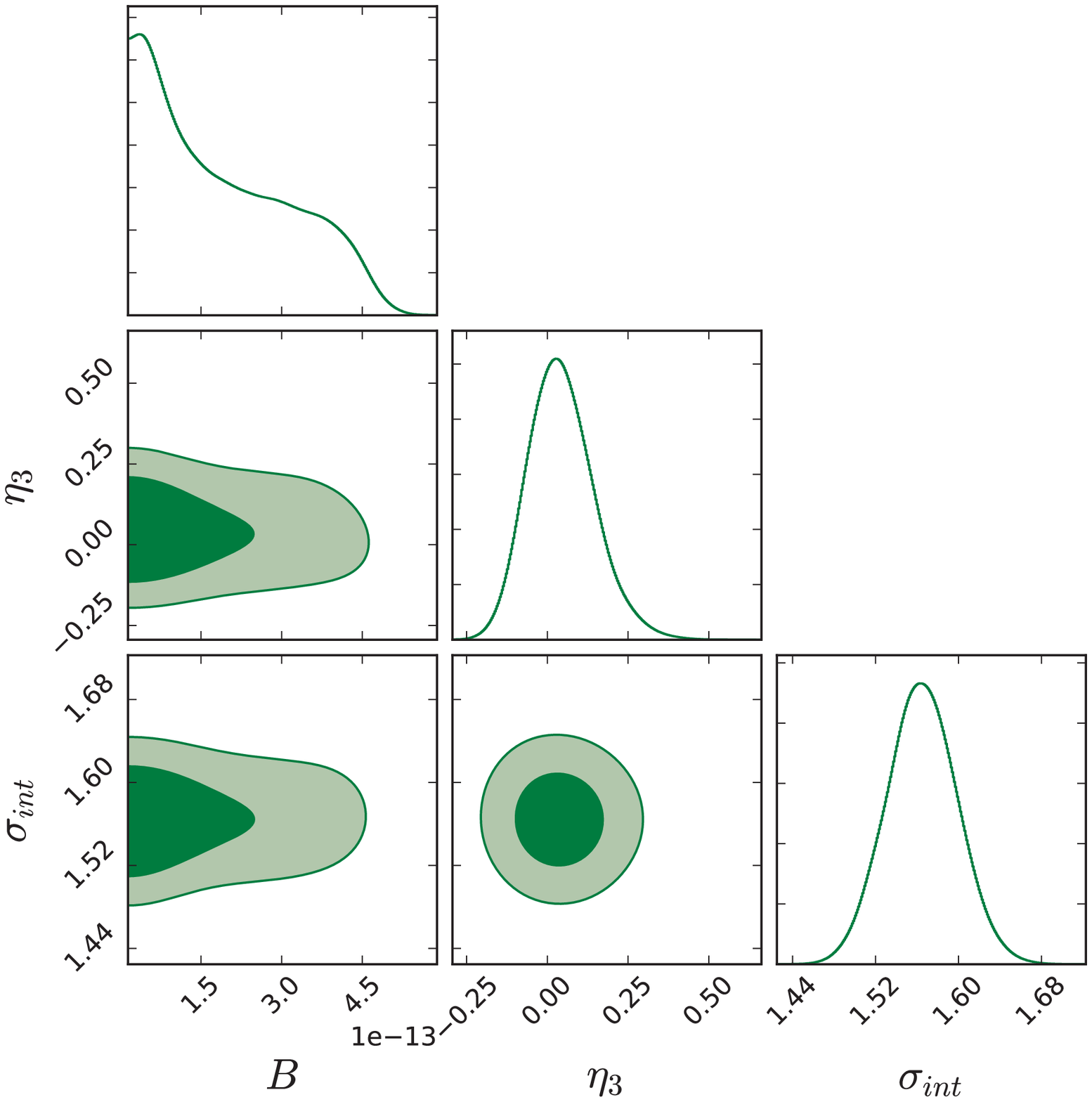}
\caption{\label{figDDR} Two dimensional and marginalized
distributions of the nuisance parameter $B$, the CDDR validity
parameter $\eta$ and the intrinsic scatter $\sigma_{int}$ in four
parameterizations of $\eta(z)$ according to Eq.~(\ref{eq:DDReta}).
 }
\end{center}
\end{figure*}

\begin{figure*}[htbp]
\begin{center}
\centering
\includegraphics[angle=0,width=70mm]{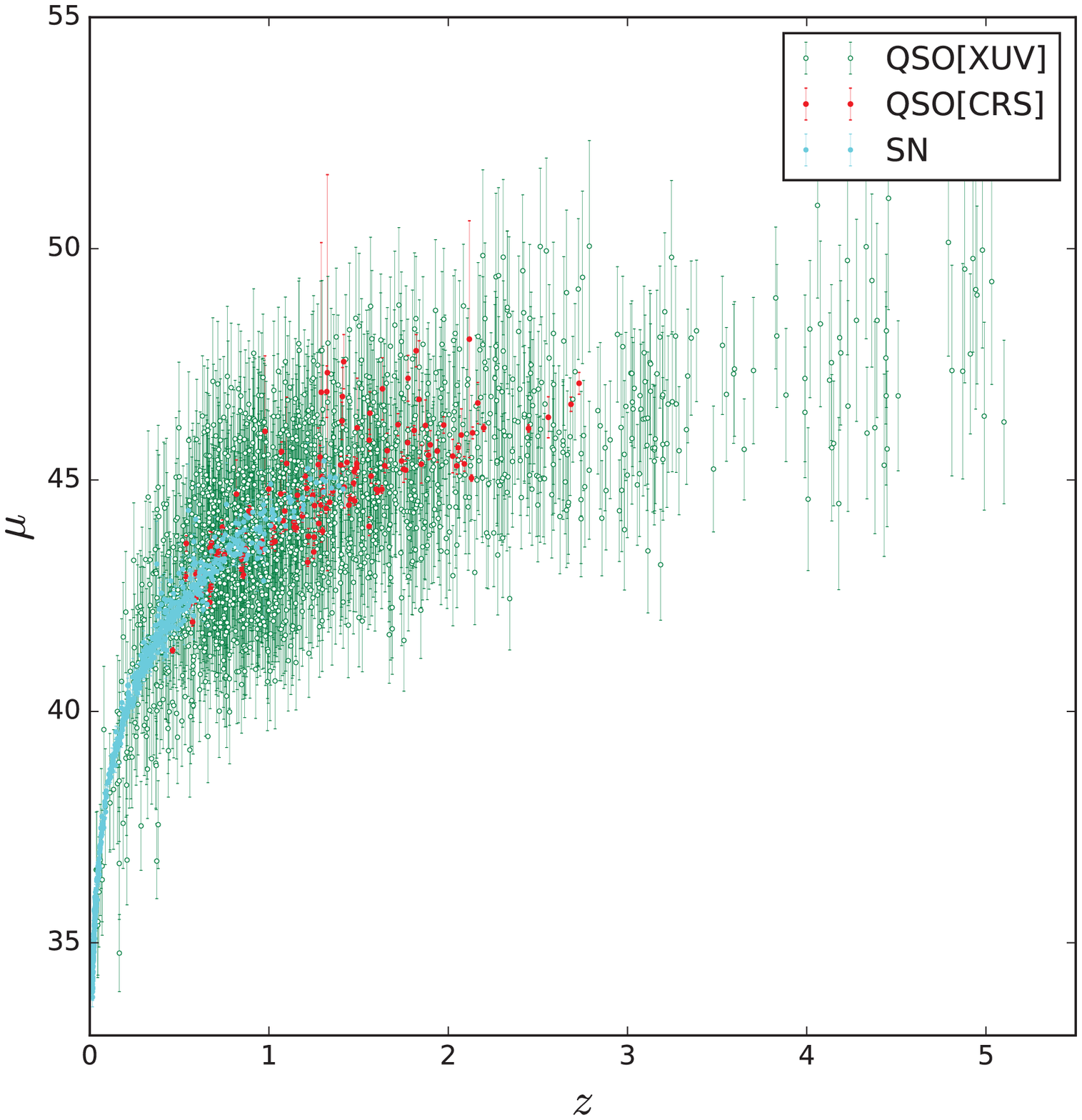}
\includegraphics[angle=0,width=63mm]{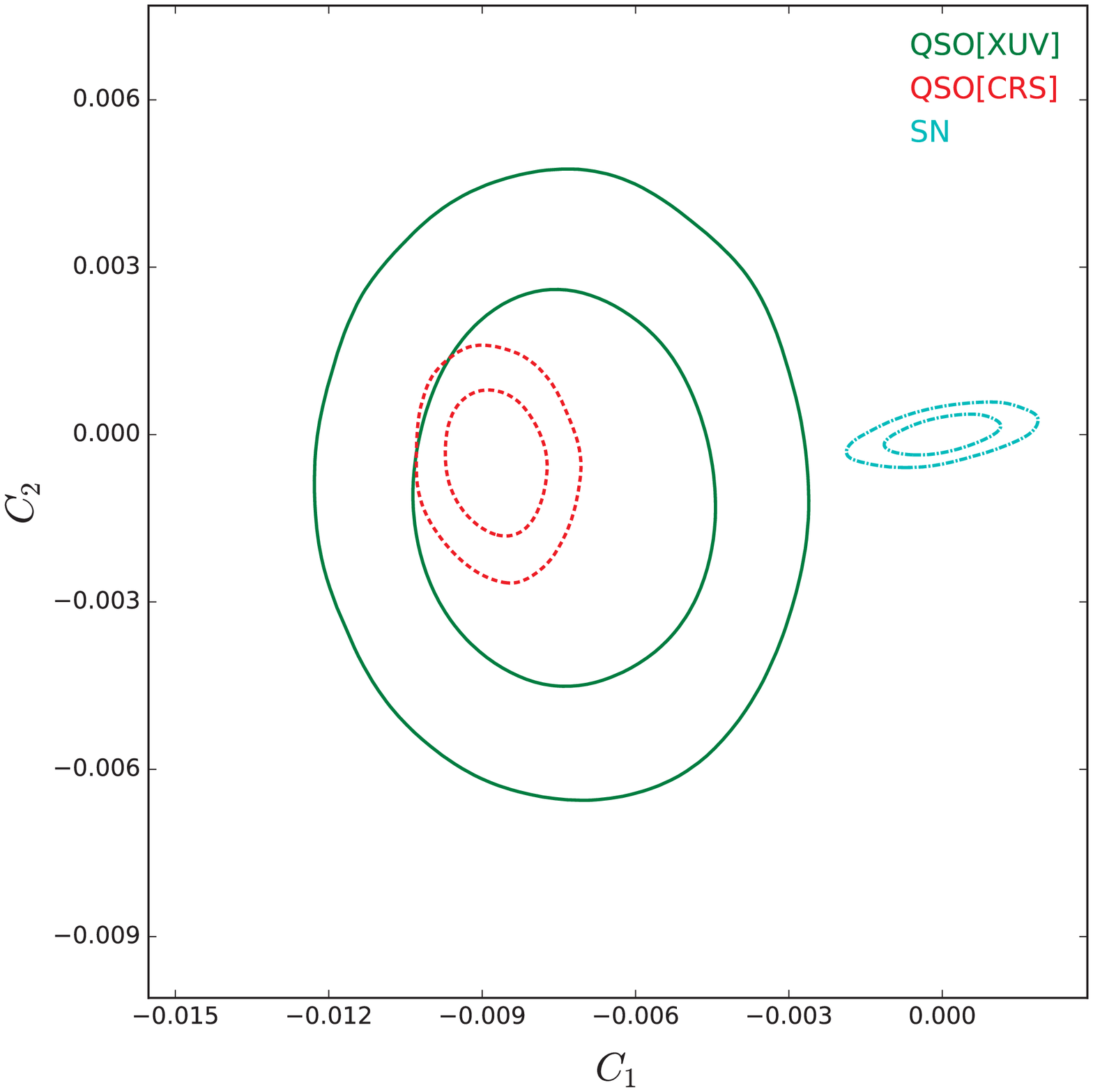}
\caption{\label{figCS} Distance modulus and confidence contours of
the crossing statistics hyperparameters based on Union2.1 SN,
QSO[XUV] and QSO[CRS] data compilations. The QSO[CRS] linear size
parameter $l=11.42\pm0.28\;pc$ and the QSO[XUV] intercept parameter
$\beta=8.24\pm0.02$ were used to get corresponding distance modulus
and both of them were calibrated from the same SN compilation
(Union2.1). }
\end{center}
\end{figure*}

Now the observed CDDR parameter $\eta_{obs}(z)$ can be expressed as
\begin{eqnarray}\label{eq:etaobs}
\nonumber \eta_{obs}(z)&=&\frac{D_L^{QSO[XUV]}(z)}{D_A^{QSO[CRS]}(z)(1+z)^2}\\
&=&B(1+z)^{-2}\theta(z)10^{(\log F_X-\gamma \log
F_{UV})/(2\gamma-2)}
\end{eqnarray}
where $B=[l_m\; 10^{[\log(4\pi)/2+\beta/(2\gamma-2)]}]^{-1}$ is the
nuisance parameter containing both the linear size scaling factor
$l_m$ and the intercept $\beta$. Concerning the uncertainty budget,
we considered uncertainties of $\theta(z)$, $\log F_{X}$, $\gamma$,
and an intrinsic scatter $\sigma_{int}$ which is meant to capture
the global intrinsic dispersion $\delta$ in QSO[XUV] data and other
unconsidered uncertainties. Intrinsic scatter was treated as a free
parameter. The uncertainty of ${\log F_{UV}}$ is negligible compared
to the uncertainty of ${ \sigma_{\log F_X}}$ hence we ignored it. So
the total uncertainty of $\eta(z)$ can be expressed as
\begin{equation}\label{eq:etaobssig}
\sigma_{\eta} =\sqrt{\sigma_{\theta}^2+\sigma_{\gamma}^2+\sigma_{\rm
logF_{X}}^2+\sigma_{int}^2}
\end{equation}
where the subsequent components are following:
\begin{eqnarray}\label{eq:etaobssig_s}
\sigma_{\theta} &= &\frac{B}{(1+z)^2}{10^{(\log F_X-\gamma \log
F_{UV})/(2\gamma-2)}}d{\theta(z)}\nonumber
\end{eqnarray}
\begin{eqnarray}
\nonumber \sigma_{\gamma} &=&\frac{B}{(1+z)^2}\theta(z){10^{(\log
F_X - \gamma \log F_{UV})/(2\gamma-2)}}\nonumber\\
&\times&{\ln 10}\frac{\log F_{UV} - \log
F_{X}}{2(\gamma-1)^2}d{\gamma}\nonumber
\end{eqnarray}
\begin{eqnarray}
\sigma_{\log F_{X}} &=&\frac{B}{(1+z)^2}\theta(z){10^{(\log F_X-\gamma \log F_{UV})/(2\gamma-2)}}\nonumber\\
&\times&{\ln 10}\frac{1}{2(\gamma-1)}d{\log F_X}
\end{eqnarray}
Free parameters in our calculations included: the nuisance parameter
$B$ (in units of $10^{-13}$), the CDDR parameters $\eta_i$ ($i = 0,
... ,3$) corresponding to four parametrizations of
Eq.~\ref{eq:DDReta} and the intrinsic scatter parameter
$\sigma_{int}$. We fitted the free parameters by maximizing the
likelihood function
\begin{equation}
\mathcal{L}({\bf{p}})=\prod\frac{1}{\sqrt{2\pi}\sigma_\eta}{\rm
exp}\left[-\frac{1}{2}\frac{(\eta_{obs}-\eta_{th})^2}{\sigma_\eta^2}\right]
\end{equation}
using the Python package of $\bf{emcee}$ \textbf{\citep{Foreman13}}
to do the Markov Chain Monte Carlo analysis.  We assumed the
following uniform priors for these parameters: $P(B)=U[0,1]$,
$P(\eta_0)=U[-1,1]$, $\sigma_{int}=U[0,2]$.

\section{Results and Discussion}\label{sec:result}

The best fitted parameter values and corresponding $1\sigma$
uncertainties are listed in Table~.\ref{tableDDR} and shown in
Fig.~\ref{figDDR}. One can see that there is no evidence for the
CDDR violation in none of the parameterizations considered. Hence
the conclusion of CDDR validity seems robust. This is consistent
with the conclusions of other works including the comparison between
luminosity distances derived form SN Ia and angular diameter
distances to galaxy clusters \citep{Holanda10,Li11}, gas mass
fraction of clusters \citep{Goncalves15}, Baryon Acoustic
Oscillations \citep{Wu15} at low redshifts, strong gravitational
lensing systems \citep{Liao16}, gamma-ray bursts \citep{Holanda17}
or compact radio sources \citep{LiX18}. Recently some authors made
forecasts concerning CDDR testing based on the simulated luminosity
distances of standard sirens detectable in the future by the
Einstein Telescope combined with strong gravitational lensing
systems \citep{Yang19} or simulated compact radio quasars
\citep{Qi19}. Assuming $\eta(z)=1+\eta_0z$ parametrization they
shown that  the 1$\sigma$ uncertainty level can reduced to 0.035 and
0.0093, respectively. Using a single population of objects,
\cite{Holanda12} tested the cosmic distance duality at low redshift
($0.14<z<0.89$) according to the measurements of the gas mass
fraction of galaxy clusters from Sunyaev-Zeldovich and X-ray surface
brightness. They obtained the following results:
$\eta_0=-0.06\pm0.16$ and $\eta_2=-0.07\pm0.24$ after excluding
objects with questionable reduced $\chi^2$. Comparing to other
independent approaches mentioned above, our method is competitive
and our results support the validity of CDDR. Underlying our
approach is the use of the same kind of objects -- quasars --
visible at high redshifts. This might alleviate some systematics
coming from different physical properties of different populations
of objects. Even though the accuracy of our results is poorer than
that forecasted from the simulated data, yet our method provided
constraints on CDDR violation more stringent than other currently
available results based on real observational data.

The linear size parameter $l_m$ and the intercept $\beta$ are hard
to determine precisely due to, respectively: the ambiguous
interpretation of the compact structure size in radio quasars and
the variation of slope in X-UV luminosity relation of quasars.
Therefore, it is necessary to calibrate the values of $l_m$ and
$\beta$ separately before using them to investigate cosmological
parameters. The linear size parameter $l_m$ was previously
calibrated with Type Ia SN \citep{Cao17b} and with Hubble parameters
\citep{Cao17a}, while the calibration of $\beta$ was discussed by
\cite{Risaliti15}. Being focused on the CCDR parameters, we
entangled $l_m$ and $\beta$ in a single nuisance parameter $B$.
However, we have checked the influence of $l_m$ and $\beta$ on the
CDDR parameters using the priors based on the above mentioned
calibrations. It turned out that it very slightly modified the
results presented in this paper. The advantage of the approach we
taken here (i.e. using a nuisance parameter $B$) was that we avoided
external calibrations which might be cosmological model dependent
and introduce tacit assumptions leading to circularity of reasoning.

Our result, i.e. confirmation of CDDR validity suggest to use it as
an assumption and discuss the consistency between distances derived
from QSO[XUV] and QSO[CRS]. The proper framework for this purpose is
set by the Crossing Statistics approach as introduced in
\citep{Shafieloo13} (for more detailed description of the method see
the references therein). Instead of smooth reconstructed distances
we will use distance moduli $\mu_{smooth}$ derived from them.
Following \citep{Shafieloo13} we represent the Crossing function by
a second order Chebyshev polynomial
\begin{equation}\label{eq:CS}
F_{II}(C_1,C_2,z)=1+C_1(\frac{z}{z_{max}})+C_2[2(\frac{z}{z_{max}})^2-1]
\end{equation}
where $C_1$, $C_2$ are the hyperparameters and $z_{max}$ is the
maximum redshift in the data compilations. Then, we fit the data to
the functions of $\mu_{smooth}^{F_{II}}(z)=\mu_{smooth}\times
F_{II}(C_1,C_2,z)$ based on $\chi^2$ statistics. For comparison, we
first use Union2.1 SN Ia data \citep{Suzuki12} to reconstruct
$\mu_{smooth}$ using Gaussian processes and then use this function
to fit SN data itself, QSO[XUV] data and QSO[CRS] data. The QSO[CRS]
linear size parameter $l=11.42\pm0.28\;pc$ and the QSO[XUV]
intercept parameter $\beta=8.24\pm0.02$ were used to derive
corresponding distances   and both of them were calibrated from the
same supernovae compilation (Union2.1). The final results of the
hyperparameters fitting are shown in Fig.\ref{figCS}. The confidence
contours of $C_1^{SN}$ and $C_2^{SN}$ is centered around the (0,0)
point as expected because the prior of the smooth mean function is
derived from supernovae data itself. According to the same smooth
mean function, the confidence contours of ($C_1^{QSO[CRS]}$,
$C_2^{QSO[CRS]}$) and ($C_1^{QSO[XUV]}$, $C_2^{QSO[XUV]}$) turn out
inconsistent with ($C_1^{SN}$, $C_2^{SN}$) while having a good
overlap region with each other. Inconsistency between supernova and
QSO[CRS](or QSO[XUV]) data suggest that it might be some systematics
in these data not properly accounted for. Moreover, this
inconsistency implies that the cosmological models best fitted to
these data compilations may be different. This issue is very
important and will be investigated in a future work.  Let us note
that the QSO[CRS] and QSO[XUV] data, both having large intrinsic
dispersions are hard to smooth directly except by using the redshift
binned data and this process may result with systematics and impact
the result. It is remarkable that the confidence contours of
QSO[CRS] and QSO[XUV] are consistent with each other which further
supports the validity of our method.

\begin{table}[htp]
\begin{center}
\caption{Results of CDDR test using four parameterizations according
to Eq.~(\ref{eq:DDReta}). $B$ is the nuisance parameter containing
both the linear size $l_m$ of QSO[CRS] standard rulers and the
intercept $\beta$ for QSO[XUV] standard candles, $\eta_i$ is the
CDDR parameter and $\sigma_{int}$ is the intrinsic scatter.}
{%{\scriptsize%\footnotesize\tiny
 \begin{tabular}{l c c c} \hline\hline
    ${\rm\eta(z)}$ & $B\times10^{-13}$ & $\eta_i$ & $\sigma_{int}$ \\ \cline{1-4} \\
 ${\rm 1+\eta_0 z}$ & $1.61^{+1.97}_{-1.12}$ & $0.01^{+0.04}_{-0.04}$ & $1.56^{+0.03}_{-0.03}$ \\
 ${\rm (1+z)^{\eta_1}}$ & $1.66^{+1.84}_{-1.11}$ & $0.02^{+0.09}_{-0.09}$ & $1.57^{+0.03}_{-0.03}$ \\
 ${\rm 1+\eta_{2}z/(1+z)}$ & $1.48^{+1.74}_{-1.02}$ & $0.13^{+0.26}_{-0.21}$ & $1.56^{+0.03}_{-0.03}$ \\
  ${\rm 1+\eta_{3}ln(1+z)}$ & $1.65^{+1.84}_{-1.15}$ & $0.04^{+0.10}_{-0.09}$ & $1.57^{+0.03}_{-0.03}$ \\
 \hline \hline
\end{tabular} \label{tableDDR}}
\end{center}
\end{table}

\section{Conclusion}\label{sec:con}

As a fundamental relation rooted in the very ground of modern
cosmology, i.e., the validity of General Relativity (or more general
-- the metric theory of gravity), the CDDR is very successful in
explaining many observational facts concerning our Universe
\citep{Planck18}. However, its slight violations might also signal
non-conservation of photon number on the way from the source to the
observer. Such effect could be due to photons decaying to axions or
due to less exotic causes like absorption by intergalactic medium.
Therefore it is very important to test it on available observational
material. So far it has been tested mainly at lower redshifts where
the observational measurements are abundant and it started to be
extended to  higher redshifts \citep{Liao16,Holanda17}. Interesting
approach is to use for this purpose the gravitational wave
measurements of coalescing compact binaries  \citep{Yang19,Qi19}.
However, we should wait for the future third generation GW detectors
until the statistics and the redshift coverage will be sufficient to
get competitive results.

In this paper, we proposed to test the CDDR using high redshift
quasars which can provide both luminosity distances and the angular
diameter distances. The angular diameter distances we used come from
the angular size measurements of compact structures in
intermediate-luminosity radio quasars \citep{Cao17a} and the
luminosity distances come from the non-linear relation between the
UV and X-ray emission \citep{Risaliti19}. We used four different
cosmic distance duality relation parameterizations: linear in
redshift $\eta(z)=1+\eta_{0} z$, power-law $\eta(z)=(1+z)^{\eta_1}$,
linear in scale factor $\eta(z)=1+\eta_{2}\; z/(1+z)$, and
logarithmic $\eta(z)=1+\eta_{3}\;{\ln}(1+z)$. The linear size
parameter $l_m$ of compact structure and the intercept parameter
$\beta$ in X-UV relation should in principle be calibrated
independently. However, it is currently very hard to achieve due to
lack of solid physical understanding of the X-UV relation and why
the intermediate-luminosity quasars could be standardizable rulers.
Moreover, external calibrators might introduce other systematics and
hidden assumptions influencing the results. In order to circumvent
these problems, we entangled both $l_m$ and $\beta$ parameter into
one external parameter $B$, which was furthermore treated as a free
parameter to fit and marginalized over. Our results, which support
the validity of the CDDR at $z\sim 3$, turned out to be more
competitive than other approaches. Given the wealth of available
high-$z$ quasars in the future, we may be optimistic about detecting
possible deviation from the CDDR within our observational volume.
Such accurate model-independent measurements of the CDDR can become
a milestone in precision cosmology.

\vspace{0.5cm}

This work was supported by the National Key R\&D Program of China
No. 2017YFA0402600; the National Natural Science Foundation of China
under grant Nos. 11690023, 11633001 and 11920101003; the Strategic
Priority Research Program of the Chinese Academy of Sciences, grant
No. XDB23000000; Beijing Talents Fund of Organization Department of
Beijing Municipal Committee of the CPC; the Interdiscipline Research
Funds of Beijing Normal University; and the Opening Project of Key
Laboratory of Computational Astrophysics, National Astronomical
Observatories, Chinese Academy of Sciences. This work was performed
in part at Aspen Center for Physics, which is supported by National
Science Foundation grant PHY-1607611. This work was partially
supported by a grant from the Simons Foundation. M.B. is grateful
for this support.

\end{document}